\documentclass{article}

\usepackage{arxiv}

\usepackage[utf8]{inputenc} 
\usepackage[T1]{fontenc}    
\usepackage{url}            
\usepackage{booktabs}       
\usepackage{amsfonts}       
\usepackage{amsmath}
\usepackage{nicefrac}       
\usepackage{microtype}      
\usepackage{xcolor}         
\usepackage{wrapfig}
\usepackage{hyperref}       
\hypersetup{
    colorlinks=true,
    linkcolor=blue,
    filecolor=magenta,      
    urlcolor=cyan,
    pdftitle={Overleaf Example},
    pdfpagemode=FullScreen,
    }

\usepackage[pdftex]{graphicx}
\graphicspath{{../figures/pdf/}{../figures/jpeg/}}
\DeclareGraphicsExtensions{.pdf,.jpeg,.png}

\title{Neural Fields for Fast and Scalable Interpolation of Geophysical Ocean Variables} 

\author{%
  J. Emmanuel Johnson\\
  Univ. Grenoble Alpes, CNRS UMR IGE, Grenoble, France\\
  \texttt{johnsonj@univ-grenoble-alpes.fr}\\
  \And
  Redouane Lguensat\\
  Institut Pierre Simon Laplace, IRD, Sorbonne Université, Paris, France \\
  \And
  Ronan Fablet\\
  IMT Atlantique, CNRS UMR Lab-STICC, Brest, France
  \AND
  Emmanuel Cosme\\
  Univ. Grenoble Alpes, CNRS UMR IGE, Grenoble, France\\
  \AND
  Julien Le Sommer\\
  Univ. Grenoble Alpes, CNRS UMR IGE, Grenoble, France\\
}

\begin{document}

\maketitle

\begin{abstract}
Optimal Interpolation (OI) is a widely used, highly trusted algorithm for interpolation and reconstruction problems in geosciences. With the influx of more satellite missions, we have access to more and more observations and it is becoming more pertinent to take advantage of these observations in applications such as forecasting and reanalysis. With the increase of the volume of available data, scalability remains an issue for standard OI and it prevents many practitioners from effectively and efficiently taking advantage of these large sums of data to learn the model hyperparameters. 
In this work, we leverage recent advances in Neural Fields (NerFs) as an alternative to the OI framework where we show 
how they can be easily applied to standard reconstruction problems in physical oceanography. We illustrate the relevance of NerFs for gap-filling of sparse measurements of sea surface height (SSH) via satellite altimetry and demonstrate how NerFs are scalable with comparable results to the standard OI. We find that NerFs are a practical set of methods that can be readily applied to geoscience interpolation problems and we anticipate a wider adoption in the future.
\end{abstract}

\section{Introduction}
\label{sec:intro}

Maps of sea surface height (SSH) provide a crucial source of information for the oceanographic community. They are used in a number of applications including research involving mesoscale ocean dynamics~\cite{SSHMesoscale,SSHEDDIES} and their role in global climate change~\cite{SSHCLIMATE,SSHTransport}. A good characterization could even lead to a better quantification of the transport of biogeochemical components~\cite{SSHTransport}. Our understanding of SSH characteristics has gotten better not only through our improved understanding of the underlying physics, but also due to the fact that we have more available observations with the introduction of satellite altimetry~\cite{SSHAltimetry, SSHAltimetry2}. However, we only ever have partial observations of SSH through satellite altimetry, so the community has put a lot of research effort into methods that can help interpolate the missing observations for use in subsequent SSH mapping products~\cite{DUACSNEW,SSHOI}.

There have been a number of different algorithmic developments that have had success in sea surface height reconstruction~\cite{SSHEOF,SSH4DVARNET}. 
In practice, the most notable of algorithms is the DUACS framework~\cite{DUACs} which is a carefully handcrafted solution based on Optimal Interpolation (OI). Developed in 1997, this framework is capable of producing global daily maps of an effective resolution of 100-200km~\cite{SWOTres,SSHOI}. This single method is solely responsible for producing many Copernicus products\footnote{DOI:\href{https://data.marine.copernicus.eu/product/SEALEVEL_GLO_PHY_L3_NRT_OBSERVATIONS_008_044/description}{10.48670/moi-00147}} such as global maps of sea surface height and sea surface temperature~\cite{DUACSNEW,SSTOI,SSTCOPERNICUS,SSTCLIMA}. However, a major downside of the OI framework is that it is computationally expensive with a very high memory consumption even at a GPU scale. This makes it very impractical for training and inference at very large scale especially with the onset of more and more available observation data~\cite{SWOT}.


In this work, we implement an alternative solution to address the computational limitations of OI: neural fields (NerF)~\cite{NeuralFields}. NerFs are a family of coordinate-based, neural network algorithms which map a set of spatial-temporal coordinates, $\mathbf{x}_\phi$, to a scalar or vector, $\mathbf{y}$, quantity of interest. These method are parametric versions of the standard OI formulation where the parameters are \textit{learned from data}. This is only possible due to the copious amounts of data we now have given the introduction of more altimetry satellites~\cite{SWOT,SWOTres}. We demonstrate that NerFs are a viable family of methods that can interpolate the SSH field given partial observations while also being orders of magnitudes faster and memory efficient than the standard OI solution. All of the preprocessing steps, machine learning modules and training regimes used in this submission is available on \textit{Github}\footnote{\url{https://github.com/jejjohnson/ml4ssh}}.

\section{Methods}
\label{sec:methods}

\subsection{Data}
\label{sec:methods:data}

In this work, we are looking at sea surface height as our observed variable of interest. The observation data is recovered from the NADIR track satellite altimetry. In particular, we parsed all data over the Gulf stream area during the time period of 01-01-2017 until 31-12-2017. We get an aggregated total of 7 altimetry tracks from the SARAL/Altika, Jason 2, Jason 3, Sentinel 3A, Cryosat-2, and Haiyang-2A satellite altimeters which is roughly $\sim$1.8 million observations in total. The Cryosat-2 altimeter data was left out as an independent validation set. We extract the spatial coordinates (latitude, longitude) and the time coordinates (seconds) as the inputs, $\mathbf{x}_\phi$, to our model and the corresponding SSH values as our outputs, $y_{\text{obs}}$. All data is available on the data challenge that has been setup on \textit{Github}\footnote{DOI:\href{https://github.com/ocean-data-challenges/2021a_SSH_mapping_OSE/}{10.5281/zenodo.5511905}}~\cite{maxime_ballarotta_2021_5511905}.

\subsection{Baselines}
\label{sec:methods:baselines}

As mentioned in the introduction, there are many methods in the literature that have been used for SSH reconstruction~\cite{SSHKRIGING,SSH4DVARNET,SSHEOF,SSHDINEOF,SSHDINEOFMulti,SSHDINAE}. However, the most prominent method that is foundational for the Copernicus products is the DUACS algorithm~\cite{DUACSNEW}. The DUACS algorithm is an OI-based scheme which tunes the kernel hyperparameters offline in a physically consistent manner with external altimetry validation datasets. While the standard OI algorithm is an excellent and trusted method for interpolation, this method suffers from scaling issues (training and inference) which can limit its use in practice. The training has a time complexity of $\mathcal{O}(N^3)$ and memory complexity of $\mathcal{O}(N^2)$ which makes this method infeasible without the appropriate hardware~\cite{GPComplexity}.

To alleviate this issue with scalability, users apply this algorithm locally by only conditioning on the observations available in the local region of interest and interpolating the remaining observations within that region. The algorithm is run along in a patch-like manner until covering the entire globe. This mosaic of many local patches results in a global map of interpolated SSH values. The local application of the OI method has several advantages: 1) It allows the OI framework to be scaled to larger datasets because it reduces the kernel matrix inversion to only the observations within the local region and 2) it removes spurious long-range correlations that are physically inconsistent~\cite{covtaper}.  

In this work, we apply the standard OI method with preconfigured hyperparameters via the local patch-based method described above to serve as the naive baseline which closely mimics the DUACS algorithm. In addition to our naive patch-based OI implementation, we also include the publicly available SSH maps, the solution to the DUACS method, as another baseline method.

\subsection{Neural Fields}
\label{sec:methods:nerfs}

Neural Fields (NerFs) are implicit models that take coordinate-based inputs (e.g. spatial-temporal) and output a scalar value or vector field for a variable of interest (e.g. sea surface height). This family of neural networks have been very successful in many computer vision applications including image regression and 3D rendering~\cite{NeuralFields} however, their adoption in Earth sciences is still at a very early stage, especially in ocean science applications~\cite{NERFHYPER}.

Standard neural networks have problems modeling high frequency signals so there were many crucial modifications that were introduced to make them successful. The first attempt included a Fourier features transformation followed by any standard neural network, i.e. the Fourier Feature Network (FFN)~\cite{NerFFFN}. This method was successfully able to capture the high frequency signals in 2D images however the gradients were noisy or nonsensical~\cite{NerFSIREN}.  Another proposed approach is the SIREN method~\cite{NerFSIREN} which uses the sine function as an activation function within a fully connected neural network. This sinusoidal activation has been explored before~\cite{SINEACT} but the recent ingenuity is the special initialization for the weights and biases enabling the network converge much faster and avoid local minima. This method also results in calculable and relatively smooth first and second order gradients which are more representative of physical systems. This is a non-exhaustive list and there are many extensions to this including the modulated version~\cite{NerFSIRENMOD} as well as Multiplicative Filter Networks (MFN)~\cite{NerFMFN}. In this work, we chose the SIREN method for its simplicity and smooth gradients but a full comparison with other NerF variants for regression problems in physical systems would be an excellent follow-up study.
\section{Experiments}
\label{sec:exps}

\subsection{Data Preparation}
\label{sec:exps:data}


\textbf{NerFs}. We used a coordinate transformation of the spatial data from spherical coordinates (lat,lon) to Cartesian coordinates (x,y,z). 
After we aggregated the altimetry datasets, we had $\sim$1.6 million data points for training and $\sim$179K data points for validation. A final evaluation is done over an entire spatial-temporal grid that is equivalent to DUACS product. The grid resolution is 0.2 degrees spatially and 1 day temporally which results in $\sim$1 million data points for the final evaluation.  The statistical metrics are computed comparing this grid product to the left out validation altimetry dataset and we plot maps for each algorithm for a visual comparison.

\subsection{Hyperparameters}
\label{sec:exps:params}

\textbf{OI.} The OI method was trained via the local patch-based version that was outlined in the previous section.
A Radial Basis Function (RBF) kernel was used with a length scale per dimension: A length scale of 7 days for the temporal dimension and a length scale of $1$ degree ($\sim$100 km) for the latitude and longitude length scale. The noise likelihood was set to $0.05$.

\textbf{NerF.} The SIREN method was trained using the standard Adam optimizer with a base learning rate of 1e-4. 
We used a cosine annealing scheduler with a linear warm-up. We used 100 epochs for the linear warm-up with a learning rate between 0 and 1e-4 followed by 2,000 epochs for the subsequent cosine annealing with a learning rate between 1e-4 and 0. We used \textit{Weights and Biases}~\cite{wandb} to perform hyperparameters sweeps on the batch size, the number of layers and the width of layers. We used the standard mean squared error as the loss function and all training was done with one Nvidia Tesla V100 GPU.



\section{Results}
\label{sec:results}

\begin{figure}[t!]
\small
\begin{center}
\setlength{\tabcolsep}{2pt}
\begin{tabular}{ccc}
 & Observations &  \\
& \hspace{6mm} \includegraphics[width=5cm,height=4cm]{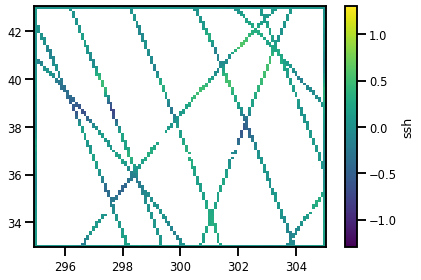} & \\
NAIVE OI & DUACS & SIREN \vspace{-4mm}\\
\includegraphics[width=4.5cm,height=4.5cm]{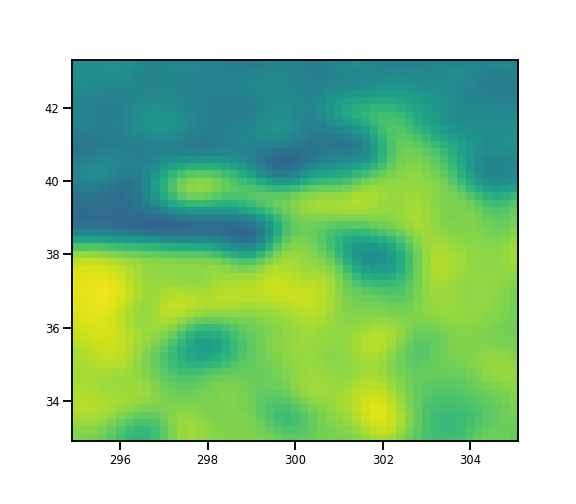} &
\includegraphics[width=4.5cm,height=4.5cm]{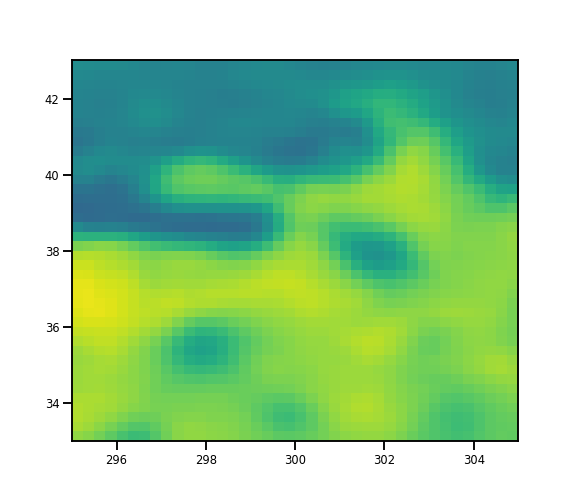} &
\includegraphics[width=5.25cm,height=4.5cm]{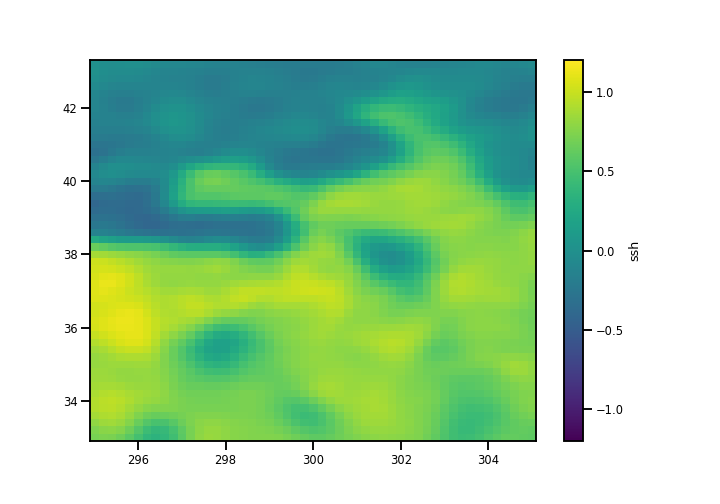} \vspace{-9mm}\\

\includegraphics[width=4.5cm,height=4.5cm]{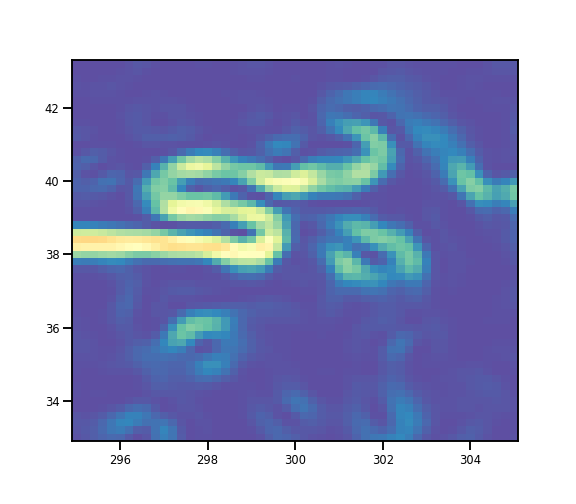} &
\includegraphics[width=4.5cm,height=4.5cm]{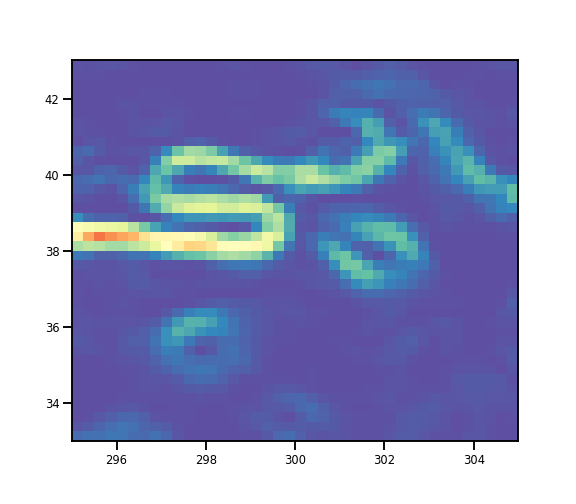} &
\includegraphics[width=5.25cm,height=4.5cm]{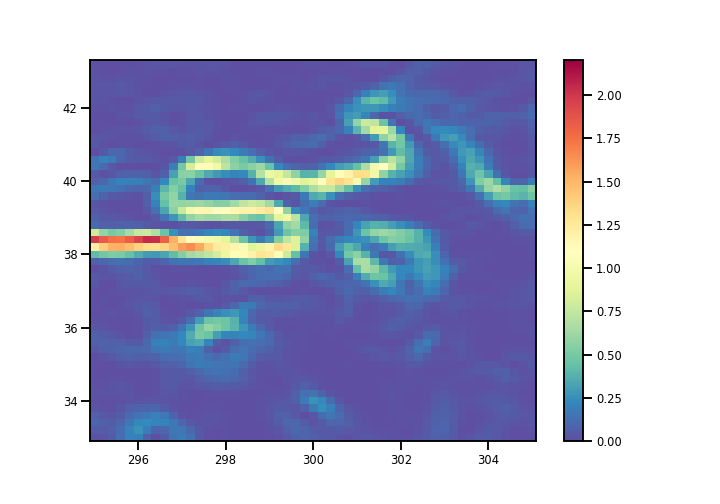} \vspace{-9mm}\\
\includegraphics[width=4.5cm,height=4.5cm]{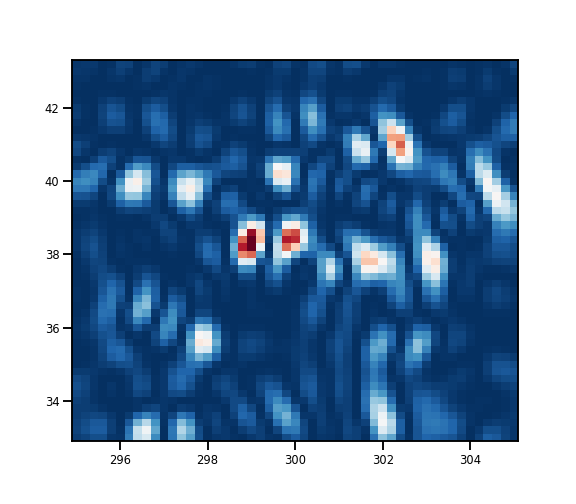} &
\includegraphics[width=4.5cm,height=4.5cm]{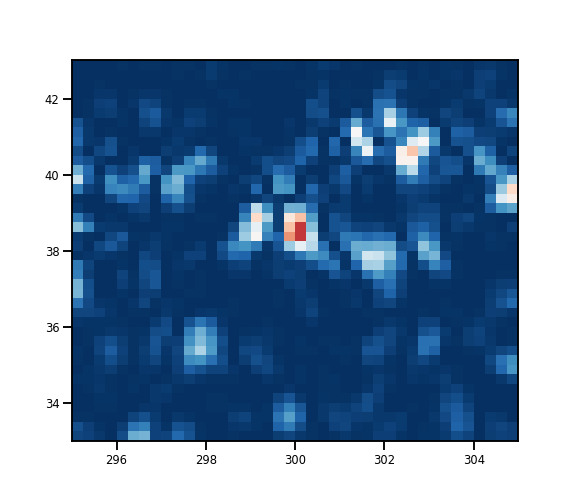} &
\includegraphics[width=5.25cm,height=4.5cm]{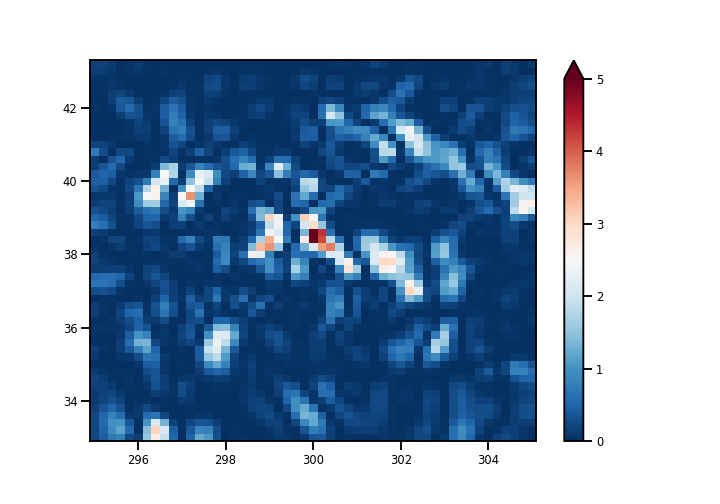} \vspace{-4mm}\\
\hspace{-5mm} \includegraphics[width=4.5cm,height=4cm]{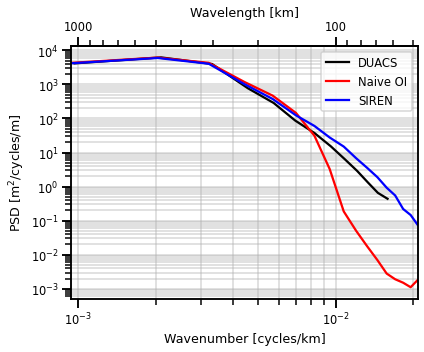} &
\hspace{-4mm} \includegraphics[width=4.5cm,height=4cm]{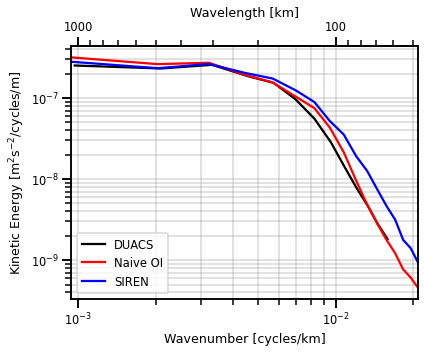} &
\hspace{-10mm} \includegraphics[width=4.5cm,height=4cm]{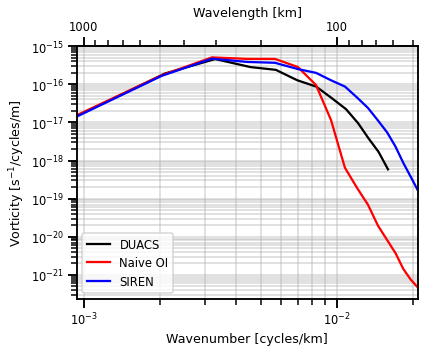}
\end{tabular}
\vspace{-4mm}
\caption{Snapshots of the daily aggregated altimetry observations (top row), the SSH field (2nd row), the norm of the gradient of SSH field (3rd row) and the Laplacian of the SSH field (4th row) over the Gulf-stream at time step 20-01-2017. Each column showcases each of the methods presented in this work: 1) Naive implementation of OI, 2) the OI in production (DUACS), and 3) the SIREN NerF. The last row showcases the spectrum for each of the variables: bottom row (left) is SSH, bottom row (middle) is the Kinetic energy and bottom row (left) is the Vorticity.}
\vspace{-5mm}
\label{fig:maps}
\end{center}
\end{figure}

\textbf{Statistics}. A snapshot of predictions, the norm of the gradients and the Laplacian can be seen in figure \ref{fig:maps}. Each of the SSH maps are very similar as SIREN method can also capture some fine scale structures like the OI methods. This is even  more evident in the Laplacian maps which result in more non-smooth structures for the SIREN method. These higher order gradients are important for other quantities of interest like the kinetic energy and vorticity and more evaluation is needed to check how each compares method compares in respective energy levels in the spectral domain. In table \ref{tb:metrics}, we see the metrics computed with the independent validation dataset; i.e. the left out Cryosat-2 altimetry track that was unseen by any model. The normalized root mean squared error (NRMSE) and effective resolution (calculated from the  signal-to-noise ratio) is comparable between all methods. For more details on the evaluation metrics, see section 3d in~\cite{SSHBFN}. Note that this is simply a local region and more evaluation is needed in other regions and at a global scale.

\textbf{Timings}. In table \ref{tb:metrics}, we see that the SIREN method is the most scalable in terms of inference\footnote{Note that we did not showcase the training times for each method because our OI is unoptimized and the DUACs methods is closed-source. However, the SIREN method is a coordinate-based method which can take a long time to train. Our model took an average of 6 hours to train not including the hyper-parameter sweeps. We recommend multi-scale training strategies and multi-node GPU architectures to get drastic speedups~\cite{NERFHYPER}.}. The naive OI framework on $\sim1$ million data points results in 1 hr on a CPU and 40 minutes on a GPU. However, the NerF algorithm predicts the same amount of data in 30 seconds on a CPU and 5 seconds on a GPU. This results in $\times120$ speedup and $\times480$ speedup respectively. This is expected due to the time complexity of matrix inversions, $\mathcal{O}(N^3)$, versus matrix multiplications, $\mathcal{O}(N)$. In addition, the OI method requires one to condition on the observations which results in an additional time and memory footprint that inhibits learning with more observational data. 

\begin{table}
  \caption{Metrics}
  \label{sample-table}
  \centering
  \begin{tabular}{llcll}
    \toprule
    Algorithm & Normalized RMSE & Effective Resolution & CPU & GPU  \\
    \midrule
    OI & $0.85 \pm 0.09$ & $140$ km & 1 hr & 40 mins   \\
    DUACs & $\mathbf{0.88 \pm 0.07}$ & $152$ km & N/A & N/A  \\
    SIREN & $0.88 \pm 0.08$ & $\mathbf{136}$ km & \textbf{30 secs} & \textbf{5 secs} \\
    \bottomrule
  \end{tabular}
  \label{tb:metrics}
 \end{table}

\section{Conclusions/Future Work}
\label{sec:conc}

\textbf{Discussion}. In this work, we introduced Neural Fields as a viable interpolation method for sea surface height. We demonstrated that we were able to match the results of the standard Optimal Interpolation algorithm by training on observations. Not only did we get results that were statistically comparable to the standard DUACS, we also achieve inference speeds and memory constraints that are viable even with a modest computational budget. Neural networks are often only good when we have a large amount of training data available. However, this work has made a case for NerFs coupled with altimetry data as a candidate algorithm for interpolation at scale. 

\textbf{Future Work}. Aside from the more detailed ablation studies related to noise and sampling density, there is much to be done to make these models more appealing and more trustworthy for the broader community. They lack immediate interpretability because they are fully parameterized machines from data. Techniques such as Explainable AI~\cite{XAI} or Physics Informed Neural Nets~\cite{NerFPINNS} could be of use to understand and constrain the NN predictions. Quantifying the predictive uncertainty is another main future direction, in fact, there are extensions that have been explored in the literature~\cite{NerFStochastic,NerFReg,NerFFlow}. Lastly, we observed that the bulk of the computational limit is training these NNs. Under a computational budget, faster training strategies~\cite{NerFScale} and better hardware would be necessary to make these methods scale to global datasets with billions of data points.

\section*{Ethical Considerations}

\label{sec:ethics}
The spatial and temporal resolution of the altimetry datasets used in this study is too coarse to affect the privacy of any individual and we only consider resolutions of $\sim100-200$km. In addition, all datasets used were publically available via the Copernicus Marine Service. However, we believe our method would be useful in higher resolution scenarios ($\sim1-50$km) which would be touching upon the territory of privacy in certain regions of the globe. The dataset itself is unlikely to enable activities that are harmful for the environment or individuals. However, if we consider that one could further develop these models for interpolating finer resolutions, one could imagine some downstream potential adverse affects on the environmental applications like overfishing or security concerns. In addition, these models are not limited to geophysical variables so one could easily apply these to land variables or human-made objects. So one should always take care when working with high spatial and temporal resolution maps.

\section*{Broader Impact}
\label{sec:impact}

The theme of interpolation is present in many applied communities with different names, e.g. kriging in ecology/hydrology, Optimal interpolation in oceanography, and Gaussian processes in statistics. We hope that this work bridges this gap between the communities and we invite other works to try to highlight concrete ways that machine learning and classic physics have commonalities.

In the oceanography community in particular, we especially hope to see more adoption of machine learning methods for interpolation. DUACS is ultimately a closed-system so the wider scientific community does not have access to the algorithm. Our OI baseline hopefully unveils some of the finer details of the method. However, in general, the standard OI methods used in the applied community cannot keep up with the massive influxes of observations we receive. So this work is a first step in demonstrating that neural networks (in particular NerFs) are a viable, simpler, and scalable alternative.

\label{sec:acks}
This work was supported by ANR (French National Research Agency, project number ANR-17- CE01-0009-01) and CNES through the SWOT Science Team program. This work was also supported by LEFE program (LEFE MANU project IA-OAC), CNES (OSTST DUACS-HR and SWOT ST DIEGO) and ANR Projects Melody (ANR-19-CE46-0011) and OceaniX (ANR-19-CHIA-0016). It benefited from HPC and GPU resources from GENCI-IDRIS (Grant 2021-101030).
We would like to thank Weights \& Biases for their support and for providing us with an academic license to use their experiment tracking tools for our work.


\bibliographystyle{plain}
\nocite{*}
\bibliography{content/biblio/ssh.bib,content/biblio/ml.bib,content/biblio/other.bib}

\end{document}